# Electric Vehicle transition in the UK

Sivapriya Mothilal Bhagavathy and Malcolm McCulloch
Energy and Power Group, Department of Engineering Science,
University of Oxford

**Abstract**

This paper provides an overview of the electric vehicle transition in the UK. The spatial disparity in the uptake of BEVs across the different regions is analysed using historic BEV sales. A forecast for future growth in BEVs (ignoring the impact of Covid-19) is performed using an s-curve model. Currently, South East England and Greater London have the highest BEV sales as a percentage of new vehicle sales. The spatial distribution of EV chargers across the different regions is also analysed. The spatial analysis clearly shows the regional disparity in the uptake of EV. South East England has the highest number of public chargers excluding Greater London. However, if we consider the number of EVs in that region, it has the second-lowest ratio of approx. 1 charger per 10 BEV. The lowest ratio being 0.8 in the West Midlands.

## 1. Introduction

In June 2019, the UK Parliament went beyond the UK's existing commitment to an 80% reduction on 1990 emissions levels by legislating for a net-zero greenhouse gas emissions target by 2050 [1]. Reflecting the different circumstances of different parts of Great Britain, in September 2019 the Scottish Parliament legislated to set a net-zero target for 2045, and the Welsh government intends to introduce legislation to set a net-zero target for Wales. Until recently, the focus has been decarbonisation of the electricity sector. However, since 2016, the transport sector is the largest emitting sector. It contributed up to 28% of the total greenhouse gas emissions in the UK as of the end of 2018 as shown in Figure 1 [2]. For the UK to achieve net-zero emissions by 2050, it is highly important to decarbonise the transport sector. The vehicle-based options for decarbonisation of the transport sector include electrification (battery electric vehicles), fuel cell vehicles (hydrogen fuel) and synthetic fuel-based internal combustion engine [3].

Electrification of vehicles is currently considered one of the options with the most potential to decarbonise the transport sector. Therefore, Electric vehicles have received a significant push from governments across the world, not just in the UK. An electric vehicle is one that uses one or more electric motors or traction motors for force. There are two types of EVs: all-electric vehicles (AEVs) and plug-in hybrid electric vehicles (PHEVs). AEVs include both battery electric vehicles (BEVs) and fuel cell electric vehicles (FCEVs). Table 1 shows the commitments to end the sales of conventional vehicles in different countries with Norway leading the path with a ban that is effective by 2025.

This paper aims to provide an overview of the spatial variation of the EV adoption and availability EV infrastructure across the UK.

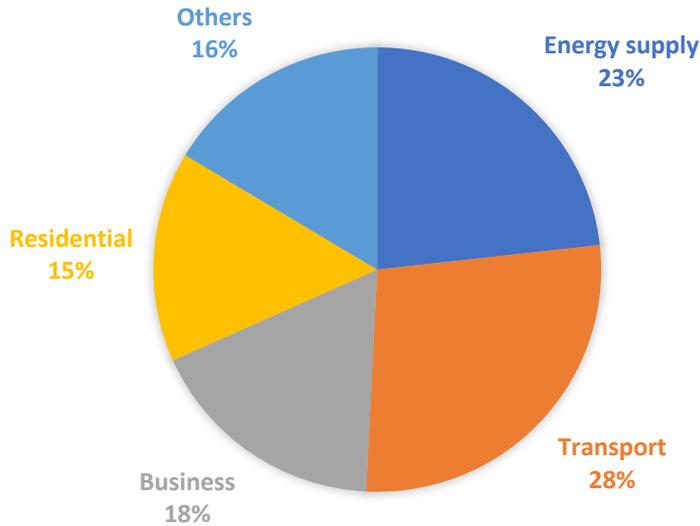

*Figure 1: Greenhouse gas emissions by source category (at the end of 2018)*

*Table 1: Commitment by different countries on a ban on the sale of conventional vehicles* [4]

| Country | Year by which ban becomes effective |
|---|---|
| Norway | 2025 |
| India, China, Slovenia, Austria, Israel, the Netherlands, Ireland | 2030 |
| Scotland | 2032 |
| UK, France, Sri Lanka, Taiwan | 2040 |

## 2. EV growth in the UK

The sale of EV in the UK has been increasing significantly over the last few years and is expected to grow exponentially in the next two decades. To promote the adoption of EVs, the UK government provided a grant of £5000 for the first 50,000 cars or until 2017 whichever was sooner [5]. This grant has since been reviewed and revised. From October 2018, only the purchase of vehicles with emissions of less than 50 g/km and a zero-emission range of at least 70 miles are eligible for grant funding, which was reduced from £4,500 to £3,500. The Department for Transport (DfT), UK and Driver and Vehicle Licensing Agency (DVLA), UK publish historic values of registered vehicles by postcode districts and battery electric vehicles by local authority [6]. Figure 2 shows the spatial distribution of BEV sales as a percentage of new vehicle sales in that region using this data. From the figure, it can be observed that South East England and Greater London has a much higher percentage of BEV sales than the rest of the country.

To forecast the growth of BEV sales, an s-curve fitting model is used. It is quite common to model technical transitions using simple s-curves [7]. The following equation describes an s-curve:

$$y(t) = \frac{1}{1+\alpha e^{-\beta t}} \quad (1)$$

where, t is the time and $\alpha, \beta$ are parameters that determine the speed and shape of transition. The limits of the equation are 0 and 1. The challenges of using an s-curve for forecasting are discussed in detail in [8]. Finding the best fit s-curve for EV growth is equivalent to finding the best fit line on a set of log axis as described in [9]. A similar approach is used to forecast the expected growth in EV sales and the resulting graphs for Oxfordshire and the UK are as shown in Figure 3. Historic quarterly data of vehicle sales and registered vehicles provided by DfT and DVLA during the period March 2011 to March 2020 is used to do the s-curve fitting. It is expected that more than 80% of the new vehicle sales in the UK will be BEVs by 2030 whereas Southeast England and London would have crossed that mark approx. five years earlier. Figure 4 shows the year at which the percentage of EV sales will be 50% of the new vehicle sales in the respective region.

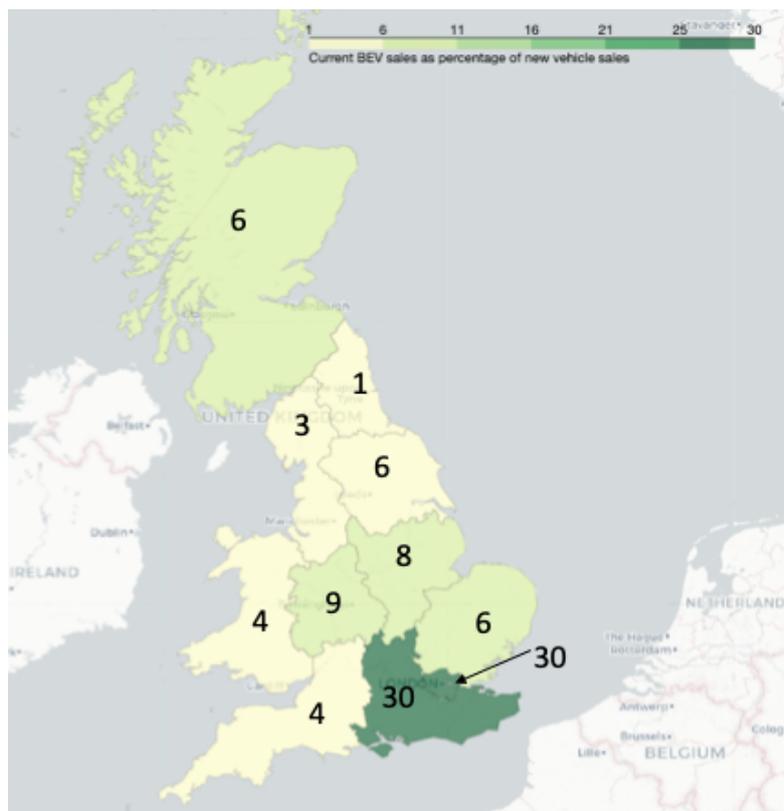

*Figure 2: Battery electric vehicle sales as a percentage of new vehicle sales at the end of March 2020 (Author's visualisation using data from the Department for Transport[6])*

From the figure, it can be observed that if the current trend continues, Southeast England and Greater London would cross the 50% mark almost a decade before Northeast England highlighting the need for further actions to promote BEVs in these regions. This also implies that the effect of EV charging will be felt in the electricity distribution networks in some regions earlier than others.

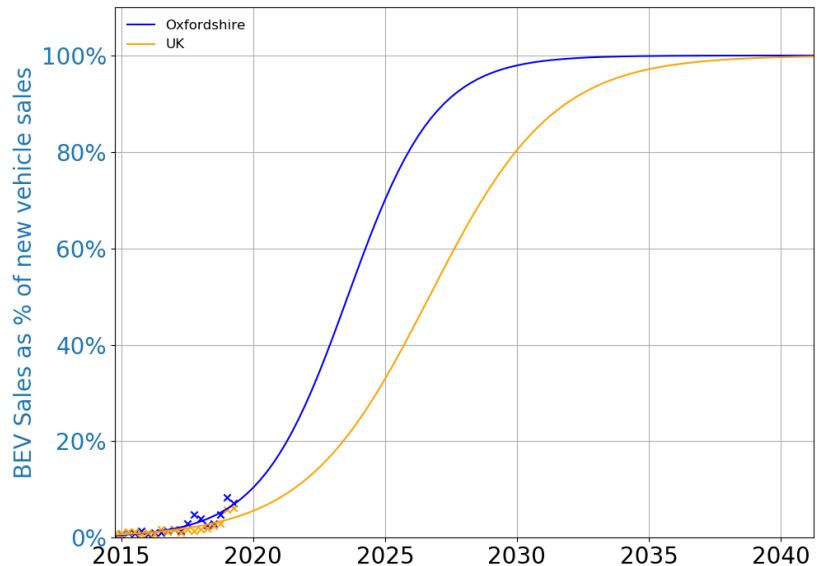

*Figure 3: Forecast for battery electric vehicle sales as a percentage of total sales in that quarter in Oxfordshire and the UK considering historic data from March 2011 to March 2020 (Data from [6])*

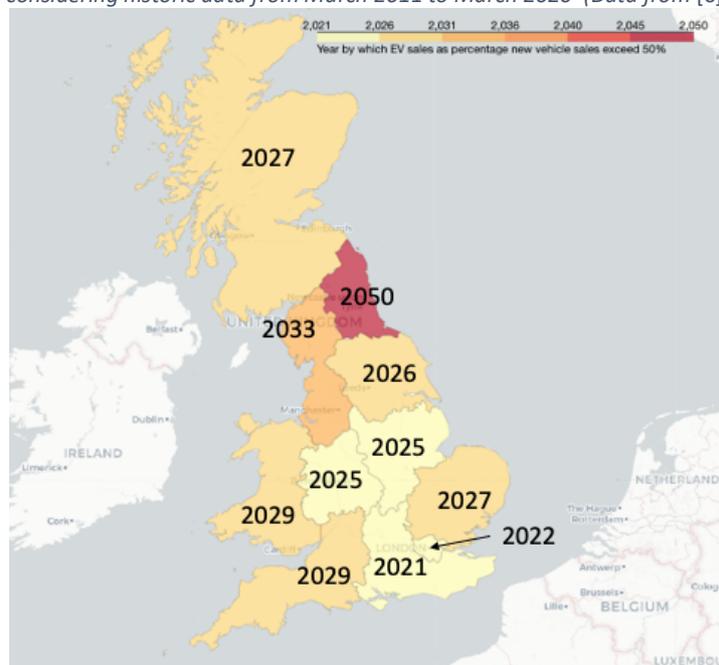

*Figure 4: Year by which battery electric vehicle sales as a percentage of total sales will cross 50% (s-curve forecast using data from March 2011 to March 2020)*

## 3. EV infrastructure growth in the UK

The overall growth in global charging infrastructure from 2013-2018 is shown in Figure 5. It can be observed that more than 90% of these are private charging points, highlighting the access to home charging amongst the early adopters of the technology. The electric vehicle home charge scheme (EVHS) provides grant funding of up to 75% towards the cost of installing electric vehicle charge points at domestic properties across the UK, but from 1st July 2019, EVHS will only support smart charge points [10]. In March 2020, this grant scheme was updated to continue for one more year albeit at a reduced grant of £350 instead of £500 per home chargers and workplace chargers. Figure 6a shows the regional distribution of total public chargers in that region. Figure 6b shows the regional distribution of the number of chargers per 10 BEVs which is a common metric used for comparison on infrastructure availability. The EU directive 2014/94/EU considers one publicly accessible charger per 10 cars as an average requirement for the facilitation of EV transition. From figure 6b it can be observed that most of the UK has achieved this average requirement. However, as the EV uptake gathers pace, the investment in infrastructure should continue to be able to provide confidence to the potential EV users.

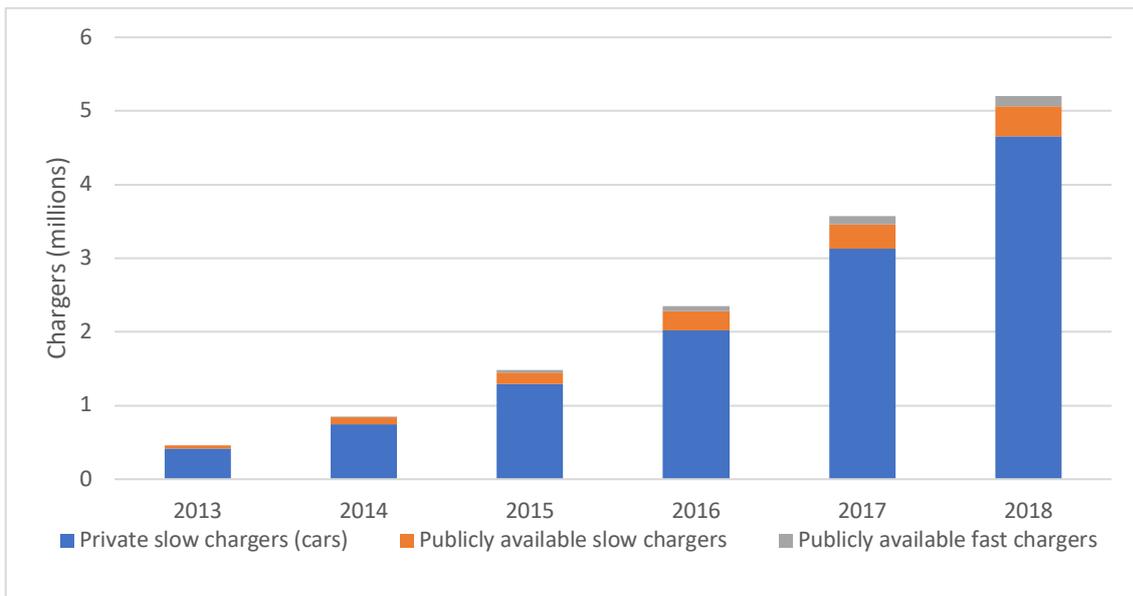

*Figure 5: Global stock of EV chargers (Data from* [11]*)*

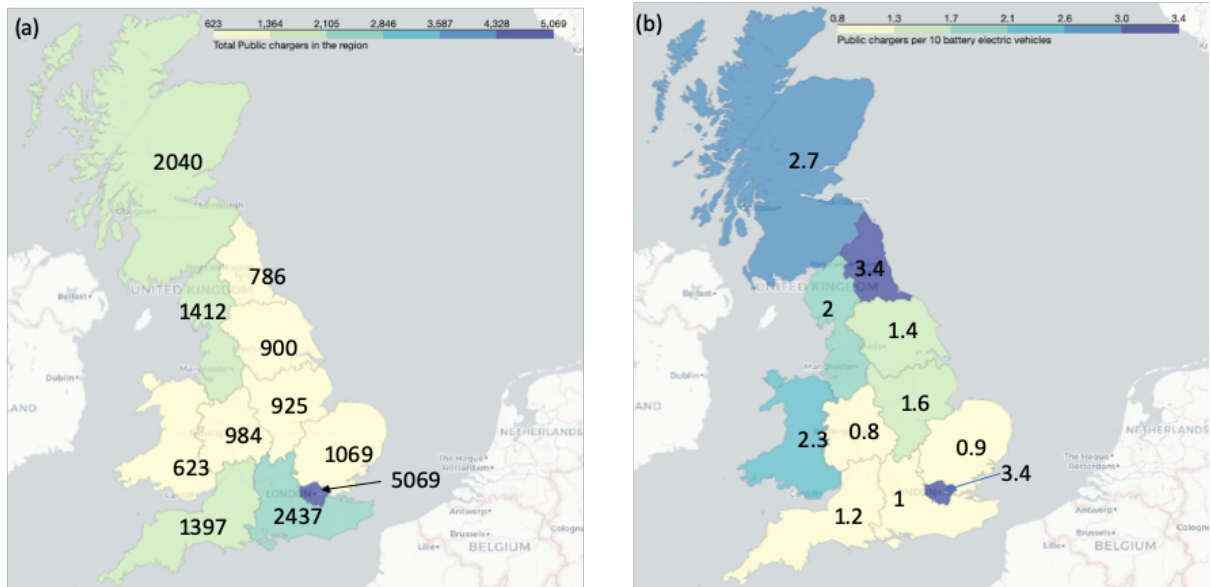

*Figure 6: (a) -Total Public chargers in the region; (b) - Public chargers per 10 BEV (Data from [12])*

## 4. Conclusions and future work

This paper showcases the status quo of the EV transition in the UK and the spatial disparity in the uptake of BEVs across the different regions. Using the historic BEV sales, a forecast for future growth in BEVs (ignoring the impact of Covid-19) is performed using an s-curve model. The spatial distribution of EV chargers across the different regions is also analysed. As the cost of EVs reduces and range of EVs increase, the inability to charge at home either due to lack of off-street parking space, multi-apartment blocks, type of tenure will be a factor inhibiting EV growth. The spatial disparity of EV growth and infrastructure helps us to evaluate the next steps in a more regional context.